\documentclass{JAC2003}
\usepackage{graphicx}
\usepackage{booktabs}

\begin{document}

\title{
ERROR EMITTANCE AND ERROR TWISS FUNCTIONS  
IN THE PROBLEM OF RECONSTRUCTION OF DIFFERENCE ORBIT PARAMETERS
BY USAGE OF BPM'S WITH FINITE RESOLUTION\vspace{-0.5cm}}

\author{V. Balandin\thanks{vladimir.balandin@desy.de}, W. Decking, N. Golubeva \\
DESY, Hamburg, Germany}

\maketitle

\begin{abstract}
The problem of errors, arising due to finite 
BPM resolution, in the difference orbit parameters, which are found as 
a least squares fit to the BPM data, is one of the standard problems
of the accelerator physics.
In this article we present a ``dynamical point of view" on this problem,
which allows us to describe properties of the BPM measurement system
in terms of the usual accelerator physics concepts of emittance and
betatron functions.
\end{abstract}

\section{INTRODUCTION}

The determination of variations in the transverse 
beam position and in the beam energy using readings of 
beam position monitors (BPMs) is one of the standard and important problems
of accelerator physics. If the optical model of the beam line and BPM 
resolutions are known, the typical choice is to let jitter parameters 
be a solution of the weighted linear least squares problem. 
Even so for the case of transversely uncoupled motion this least squares 
problem can be solved ``by hand" (see, for example ~\cite{LohseEmma, YChao}), 
the direct usage of obtained analytical solution as a tool for designing 
of a ``good measurement system" does not look to be fairly straightforward. 
It seems that a better understanding of the nature of the problem is still desirable.

A step in this direction was made in the papers ~\cite{BPM_1, BPM_2}, where
dynamic was introduced into this problem which in the beginning 
seemed to be static. When one changes the position of the reconstruction 
point, the estimate of the jitter parameters propagates along the beam
line exactly as a particle trajectory and it becomes possible (for every fixed
jitter values) to consider a virtual beam consisting from virtual particles 
obtained as a result of application of least squares reconstruction procedure 
to ``all possible values'' of BPM reading errors. The dynamics of the centroid 
of this beam coincides with the dynamics of the true difference orbit 
and the covariance matrix of the jitter reconstruction errors can be treated
as the matrix of the second central moments of this virtual beam
distribution.

In accelerator physics a beam is characterized by its emittances, energy spread, 
dispersions, betatron functions and etc. All these values immediately become the 
properties of the BPM measurement system. In this way one can compare two BPM systems
comparing their error emittances and error energy spreads, or, for a
given measurement system, one can achieve needed balance between coordinate
and momentum reconstruction errors by matching the error betatron functions 
in the point of interest to the desired values.
In this article we illustrate this dynamical point of view on the BPM measurement 
system considering the case of transversely uncoupled
nondispersive beam motion
(inclusion of the energy degree of freedom and multiple examples can be found
in cited above papers ~\cite{BPM_1, BPM_2}).
As application, we formulate requirements on the BPM measurement system
of the high-energy intra-bunch-train feedback system (IBFB) of the 
European X-Ray Free-Electron Laser (XFEL) Facility in terms 
of introduced concepts of error emittance
and error Twiss parameters ~\cite{XFEL, IBFB}. 

\section{STANDARD LEAST SQUARES SOLUTION}

We will assume that the transverse particle motion is uncoupled
in linear approximation and will use the variables 
$\,\vec{z} = (x, \, p_x)^{\top}\,$ 
for the description of the horizontal beam oscillations.
As orbit parameters we will understand  values of 
$\,x\,$ and $\,p_x\,$ given in some predefined point 
in the beam line (reconstruction point with longitudinal position $\,s = r\,$) 
and as transverse jitter in this point we will mean the difference 
$\,\delta \vec{z}(r) = \left(
x(r) - \bar{x}(r),\, p_x(r) - \bar{p}_x(r)\right)^{\top}$
between parameters of the instantaneous orbit
and parameters of some predetermined ``golden trajectory'' 
$(\bar{x},\, \bar{p}_x)^{\top}$.

Let us assume that we have $n$ BPMs in our beam line placed at positions 
$\,s_1, \ldots, s_n\,$ and they deliver readings
$\vec{b}_{c} = \left(b_1^c, \ldots, b_n^c\right)^{\top}$
for the current trajectory with previously recorded 
observations for the golden orbit being 
$\vec{b}_{g} = \left(b_1^g, \ldots, b_n^g \right)^{\top}$.
Suppose that the difference between these readings 
can be represented in the form

\noindent
\begin{eqnarray}
\delta \vec{b}_{\varsigma}
\, \stackrel{\rm def}{=} \,
\vec{b}_{c} \,-\, \vec{b}_{g} \, = \,
\left(
\begin{array}{c}
x(s_1) - \bar{x}(s_1) \\
\vdots                \\
x(s_n) - \bar{x}(s_n)
\end{array}
\right) \,+\, \vec{\,\varsigma} ,
\label{SEC1_3}
\end{eqnarray}

\noindent
where the random vector 
$ \vec{\,\varsigma} = (\varsigma_1, \ldots, \varsigma_n)^{\top}$ 
has zero mean and positive definite covariance matrix $V_{\varsigma}$.

Let $\,A_m(r)\,$ be a symplectic transfer matrix from location of the
reconstruction point to the $m$-th BPM location

\noindent
\begin{eqnarray}
A_m(r) \, = \,
\left(
\begin{array}{ll}
a_m(r) & c_m(r)\\
e_m(r) & d_m(r)
\end{array}
\right)
\label{m1}
\end{eqnarray}

\noindent
and let us assume that the Cholesky factorization 
$\,V_{\varsigma} = R_{\varsigma}^{\top} R_{\varsigma}\,$ 
of the covariance matrix $\,V_{\varsigma}\,$ is known.
As usual, we will find an estimate
$\,\delta \vec{z}_{\varsigma} (r)\,$
for the difference orbit parameters in the presence of BPM reading
errors by solving the following weighted linear least squares problem

\noindent
\begin{eqnarray}
\min_{ \delta \vec{z}_{\varsigma} }  \;
\big\| R_{\varsigma}^{-\top} \big(M \cdot \delta \vec{z}_{\varsigma} \,-\, 
\delta \vec{b}_{\varsigma} \,\big) \big\|_2^2,
\label{SEC1_9}
\end{eqnarray}

\noindent
where 

\noindent
\begin{eqnarray}
M \,=\,
\left(
\begin{array}{cc}
a_1(r) & c_1(r)\\
\vdots & \vdots\\
a_n(r) & c_n(r)
\end{array}
\right).
\label{a2}
\end{eqnarray}

\noindent
If the phase advance between at least 
two BPMs is not multiple of $180^{\circ}$, 
then the solution of the problem (\ref{SEC1_9}) is unique 
and is given by the well known formula

\noindent
\begin{eqnarray}
\delta \vec{z}_{\varsigma}(r) \, = \,
\left(M^{\top}(r) V_{\varsigma}^{-1} M(r) \right)^{-1}
M^{\top}(r) \, V_{\varsigma}^{-1} \cdot \delta \vec{\,b}_{\varsigma}.
\label{SEC1_15}
\end{eqnarray}

\noindent
The calculation of the covariance matrix of the errors 
is also standard and gives the following result

\noindent
\begin{eqnarray}
V_z(r) \, \stackrel{\rm def}{=} \,
{\cal V} \left(\, \delta \vec{z}_{\varsigma}(r) \, \right)
\,=\,
\left(M^{\top}(r) \, V_{\varsigma}^{-1} \, M(r) \right)^{-1}.
\label{SEC1_16}
\end{eqnarray}

\section{BEAM DYNAMICAL PARAMETRIZATION OF COVARIANCE MATRIX OF
RECONSTRUCTION ERRORS}

Let $\,A(r_1, \, r_2)\,$ be a matrix which transports particle coordinates
from the point with the longitudinal position  $\,s = r_1\,$ to the point
with the position  $\,s = r_2\,$. 
It is not difficult to show that for 
any given value of $\,\vec{\,\varsigma}\,$
the estimate of the difference orbit parameters  
$\,\delta \vec{z}_{\varsigma}\,$
propagates along the beam line exactly as particle trajectory

\noindent
\begin{eqnarray}
\delta \vec{z}_{\varsigma}(r_2) \; = \;
A( r_1 , \, r_2 ) \cdot \delta \vec{z}_{\varsigma}(r_1),
\label{bdp_2}
\end{eqnarray}

\noindent
as one changes the position of the reconstruction point.
So we can consider a virtual beam consisting from 
virtual particles obtained as a result of application of
formula (\ref{SEC1_15}) to ``all possible values'' of the
error vector $\,\vec{\,\varsigma}$. The dynamics of the
centroid of this virtual beam 
$\,\delta \vec{z}_0(r) = \big<\delta \vec{z}_{\varsigma}(r)\big>\,$ 
coincides with the dynamics of the true difference orbit 
$\delta \vec{z}(r)$ and the error covariance matrix (\ref{SEC1_16}) 
can be treated as the matrix of the second central moments of this virtual beam
distribution and satisfies the usual transport equation

\noindent
\begin{eqnarray}
V_z(r_2) \;=\; A( r_1 , \, r_2 ) \, V_z(r_1) \,
A^{\top}( r_1 , \, r_2 ) .
\label{bdp_3}
\end{eqnarray}

Consequently, for the description of the propagation of
the reconstruction errors along the beam line,
one can use the accelerator physics notations
and represent the error covariance matrix in the familiar form

\noindent
\begin{eqnarray}
V_z(r) =
\epsilon_{\varsigma} 
\, 
\left(
\begin{array}{rrr}
  \beta_{\varsigma}(r) & -\alpha_{\varsigma}(r)\\
-\alpha_{\varsigma}(r) &  \gamma_{\varsigma}(r)
\end{array}
\right),
\label{TwCVM_1}
\end{eqnarray}

\noindent
where $\beta_{\varsigma}(r)$ and $\alpha_{\varsigma}(r)$
are the error Twiss parameters and

\noindent
\begin{eqnarray}
\epsilon_{\varsigma} = \sqrt{\,\det V_z(r)\,}
\label{SEC2_5}
\end{eqnarray}

\noindent
is the invariant error emittance.

What is interesting about the error Twiss parameters is the fact that they 
are not simply one of many betatron functions which could propagate 
through our beam line, they are by themselves solutions of some minimization
problem. 
For simplicity of formulations let us consider the case when
readings of different BPMs are uncorrelated, i.e. when

\noindent
\begin{eqnarray}
V_{\varsigma} \,=\,
\mbox{diag}
\left(\,
\sigma_1^2 ,\, \sigma_2^2 ,\, \ldots ,\, \sigma_n^2\,
\right)\,.
\label{SEC1_17}
\end{eqnarray}

\noindent
Then, under the assumption that the phase advance between at least
two BPMs is not a multiple of $180^{\circ}$, 
the error Twiss parameters are unique solutions to the problem

\noindent
\begin{eqnarray}
\min_{\beta(r), \, \alpha(r)}
\;\;\sum \limits_{ m = 1 }^{ n } \frac{\beta(s_m)}{\sigma_m^2} 
\label{SEC2_11}
\end{eqnarray}

\noindent
and this minimum is equal to $\,2 / \epsilon_{\varsigma}$.
Besides that, 
the error betatron functions (and only they) satisfy

\noindent
\begin{eqnarray}
\sum_{ m = 1 }^{ n } \frac{\beta_{\varsigma}(s_m)}{\sigma_m^2}
\cdot
\left(
\begin{array}{c}
\cos\left(2 \mu_{\varsigma}(r, s_m)\right)\\
\sin\left(2 \mu_{\varsigma}(r, s_m)\right)
\end{array}
\right)
= 0,
\label{SEC2_16}
\end{eqnarray}

\noindent
where $\,\mu_{\varsigma}(r, s_m)\,$ is the phase advance calculated from
the point $\,s = r\,$ to the point $\,s = s_m\,$.

\section{COURANT-SNYDER INVARIANT AS ERROR ESTIMATOR}

Beam dynamical point of view on the BPM measurement system
leads us, almost unavoidably, to the introduction of the Courant-Snyder 
quadratic form as error estimator.
Let 
$\,\beta_0, \, \alpha_0, \, \gamma_0\,$ be the 
design Twiss parameters and

\noindent
\begin{eqnarray}
I_x (r, \, \vec{z}\,) = 
\gamma_0(r) \,x^{2} + 2 \alpha_0(r) \,x p_x  + \beta_0(r) \,p_x^{2} 
\label{IFB_1}
\end{eqnarray}

\noindent
the corresponding Courant-Snyder quadratic form. 
Using this quadratic form
we introduce the random variable

\noindent
\begin{eqnarray}
I_x^{\varsigma} \,=\, I_x ( r, \, \delta \vec{z}_{\varsigma}(r) -
\delta \vec{z}_0(r)) .
\label{SEC_GD_1}
\end{eqnarray}

\noindent
The mean value of this random variable is equal

\noindent
\begin{eqnarray}
\big< \, I_x^{\varsigma} \, \big> 
\,=\, 2 \,\epsilon_{\varsigma} \, m_p ,
\label{App1}
\end{eqnarray}

\noindent
where $\,m_p = m_p(\beta_{\varsigma},\,\beta_0)\,$ is the mismatch
between the error and the design betatron functions.
The right hand side in (\ref{App1}),
as it could be expected, does not depend on the position of the reconstruction 
point, but it depends not only on the error emittance but also on the 
the design and the error betatron functions.
So if one will use Courant-Snyder quadratic form for the
estimation of the properties of the BPM measurement system, 
then the figure of merit for the quality of this system will be not
the error emittance alone, but the product of the error emittance and the mismatch 
between the error and the design Twiss parameters. Large mismatch
can spoil the properties of the measurement system
even for the case when the error emittance is small.

If we will assume that the random vector $\,\vec{\,\varsigma}\,$ 
has a multivariate normal distribution, then it becomes possible to 
find not only higher order moments of the random variable $\,I_x^{\varsigma}\,$, 
but also its probability density.
This density $\,p(t)\,$ is equal to zero for negative values of its argument,
and for $\,t \geq 0\,$

\noindent
\begin{eqnarray}
p(t) \,=\, \frac{1}{2 \epsilon_{\varsigma}} \,
I_0 \left(\sqrt{m_p^2- 1} \,\, 
\frac{t}{2 \epsilon_{\varsigma}}\right)
\exp \left(-m_p\,\frac{t}{2 \epsilon_{\varsigma}}\right),
\label{probD}
\end{eqnarray}

\noindent
where $I_0$ is the modified Bessel function of zero order.

\section{STUDIES OF THE BPM RESOLUTION
REQUIRED FOR THE IBFB SYSTEM OF THE EUROPEAN XFEL}

The typical requirement for the transverse (horizontal) beam stability
at the entrance of the SASE undulator is usually formulated 
in the terms of beam sigmas and can be written in the form

\noindent
\begin{eqnarray}
I_x (r, \, \delta \vec{z}_0(r)) \;\leq \;n_{\sigma}^2 \,\varepsilon_x ,
\label{IBFB_1}
\end{eqnarray}

\noindent
where $\varepsilon_x$ is non-normalized rms emittance, 
$n_{\sigma}$ is some predefined number of beam sigmas,
and $\,\delta \vec{z}_0\,$ is the difference between 
parameters of the instantaneous and the golden trajectories.
In order to satisfy inequality (\ref{IBFB_1})
with small transverse emittances required for the
SASE FEL process and with typical limitation on $n_{\sigma}$
to be not larger than $0.1$,
the active beam stabilization system (transverse feedback)
is planned to be used at the European XFEL Facility ~\cite{XFEL,IBFB}.

The purpose of this section is to get first idea about 
BPM resolution needed for such feedback system.
Because actual feedback performance will strongly depend 
not only on BPM resolutions but also
on the interplay between the properties of the real beam jitter
and the feedback algorithm used, let us, for the first guess, 
consider very simplified idealized feedback system which
could act without delay on the same bunch which was measured
and whose kickers do not introduce own correction errors.
In this situation the only problem left is that $\,\delta \vec{z}_0\,$ 
remains unknown for us and
instead feedback BPM's deliver
us an estimate $\,\delta \vec{z}_{\varsigma}\,$, which 
includes the effect of the BPM reading errors. 
Let us write

\noindent
\begin{eqnarray}
I_x (r, \, \delta \vec{z}_0(r))  =
I_x ( r, \,[\delta \vec{z}_{\varsigma}(r) -
\delta \vec{z}_0(r)] - [\delta \vec{z}_{\varsigma}(r)]) 
\label{IBFB_2}
\end{eqnarray}

\noindent
and assume that, according to the above discussions, $\,\delta \vec{z}_{\varsigma}\,$
can be perfectly corrected by feedback kickers to zero.
Then the criteria (\ref{IBFB_1}) can be reformulated in the form

\noindent
\begin{eqnarray}
\Pr \left( I_x^{\varsigma}\,\leq \,n_{\sigma}^2 \,\varepsilon_x \right) \,\geq\,p_0,
\label{IBFB_3}
\end{eqnarray}

\noindent
where $p_0$ is some predetermined probability of correction success.
Let $\,t = t(p_0, m_p, \epsilon_{\varsigma})\,$ be solution
of the equation

\noindent
\begin{eqnarray}
\Pr \left( I_x^{\varsigma}\,\leq \,t\right) \,=\,p_0.
\label{IBFB_4}
\end{eqnarray}

\noindent
Then it can be represented in the form 
$\,t = 2 \epsilon_{\varsigma}\, a(p_0, m_p)$, where
the function $\,a(p_0, m_p)\,$ can be found from

\noindent
\begin{eqnarray}
\int\limits_0^{a(p_0, m_p)}
I_0 \left(\sqrt{m_p^2 - 1} \,\, \tau \right) 
\exp(-m_p \,\tau)\, d \tau \,=\, p_0.
\label{IBFB_6}
\end{eqnarray}

\noindent
Comparing (\ref{IBFB_3}) and (\ref{IBFB_4}) one obtains the
requirement

\noindent
\begin{eqnarray}
2 \epsilon_{\varsigma} a(p_0, m_p) \,\leq\, n_{\sigma}^2 \varepsilon_x .
\label{IBFB_7}
\end{eqnarray}

\noindent
For two uncorrelated BPMs with equal resolutions the error
emittance is 
$\epsilon_{\varsigma} = \sigma_{bpm}^2 / |r_{12}|$ and
this together with (\ref{IBFB_7}) gives

\noindent
\begin{eqnarray}
\sigma_{bpm} \,\leq\,n_{\sigma}
\sqrt{(\varepsilon_x |r_{12}|) / (2 a(p_0, m_p))}.
\label{IBFB_8}
\end{eqnarray}

\noindent
In the framework of the model considered,
the right hand side of the inequality (\ref{IBFB_8}) gives maximal allowed
BPM resolution for the case when two BPM's will be used for the measurement
of the horizontal beam jitter. If one will use the same BPM's for measuring
jitters in both transverse planes simultaneously,
then as maximal allowed resolution one has to take the minimum of the 
right hand sides of the inequality (\ref{IBFB_8}) and analogous inequality
written for the vertical beam motion. 
As an important practical example, 
Fig.1 shows the maximal allowed resolution of two feedback BPM's
in the situation when they will be used for the measurement of both
transverse jitters simultaneously for the case of the IBFB beam line
of the European XFEL Facility ~\cite{IBFB}. Note that calculations 
presented at this figure were done for $\,p_0 = 0.95$, $\,n_{\sigma} = 0.1,$
beam energy of $17.5\,GeV$ and normalized emittances of $1.4\,mm\cdot mrad$.

\begin{figure}[t]
    \centering
    \includegraphics*[width=75mm]{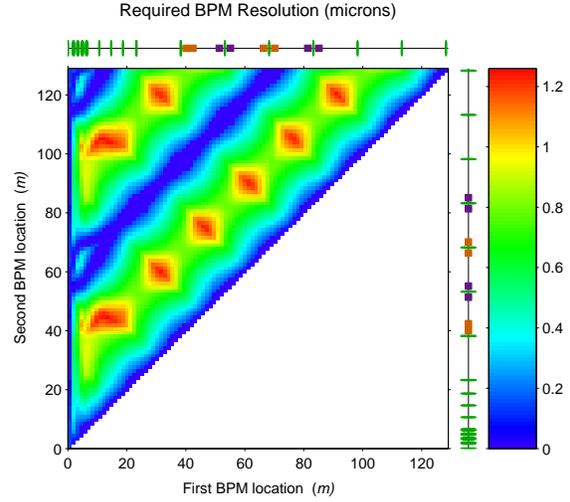}
    \caption{Maximal allowed resolution of two feedback BPM's
    as a function of their positioning in the beam line.}
    \label{figBeta}
\end{figure}

\end{document}